\documentclass[sigconf]{acmart}

\usepackage{booktabs} 
\usepackage{balance}
\usepackage{multirow}
\usepackage{adjustbox}
\usepackage{xcolor}
\usepackage{hyperref}
\usepackage{mfirstuc}
\usepackage{titlecaps}
\usepackage{stringstrings}
\usepackage{booktabs}
\usepackage{tabularx}
\usepackage{tabularx,booktabs} 

\usepackage{color}
\usepackage{url}
\usepackage{balance}
\usepackage[english]{babel}
\usepackage[utf8]{inputenc} 
\usepackage[T1]{fontenc}
\usepackage[normalem]{ulem} 
\usepackage{xspace}
\usepackage{diagbox}
\usepackage{amssymb}
\usepackage{pifont}
\usepackage{multirow}
\usepackage{rotating}
\usepackage{wrapfig}

\usepackage{amssymb}
\usepackage{amsmath}
\usepackage{graphicx}
\usepackage[ruled]{algorithm2e}
\usepackage{breqn}
\usepackage{multirow}
\usepackage{capt-of}
\usepackage[T1]{fontenc}
\usepackage{letltxmacro}
\usepackage{lipsum}

\usepackage{algorithm2e}
\usepackage{letltxmacro}
\usepackage{bookmark}

\begin{document}

\copyrightyear{2018}
\acmYear{2018}
\setcopyright{acmcopyright}
\acmConference[ICPC '18]{ICPC '18: 26th IEEE/ACM International Confernece on Program Comprehension }{May 27--28, 2018}{Gothenburg, Sweden}
\acmBooktitle{ICPC '18: ICPC '18: 26th IEEE/ACM International Confernece on Program Comprehension , May 27--28, 2018, Gothenburg, Sweden}
\acmPrice{15.00}
\acmDOI{10.1145/3196321.3196349}
\acmISBN{978-1-4503-5714-2/18/05}

\title{Identifying Software Components from Object-Oriented APIs Based on Dynamic Analysis}

\author{Anas Shatnawi}
\affiliation{%
  \institution{University of Milan-Bicocca}
  \city{Milan} 
  \country{Italy}
}
\email{anas.shatnawi@unimib.it}

\author{Hudhaifa Shatnawi}
\affiliation{
  \institution{Maharishi University of Management}
  \city{Fairfield, Iowa} 
  \country{USA}
}
\email{hshatnawi@mum.edu}

\author{Mohamed Aymen Saied}
\affiliation{%
  \institution{Concordia University}
  \city{Montreal, Quebec} 
  \country{Canada}
}
\email{med.aymen.saied@gmail.com}

\author{Zakarea	Al Shara}
\affiliation{%
  \institution{University of Montpellier}
  \city{Montpellier} 
  \country{France}
}
\email{zakarea.alshara@gmail.com}

\author{Houari Sahraoui}
\affiliation{%
  \institution{University of Montreal}
  \city{Montreal, Quebec} 
  \country{Canada}
}
\email{sahraouh@iro.umontreal.ca}

\author{Abdelhak Seriai}
\affiliation{%
  \institution{University of Montpellier}
  \streetaddress{Viale Sarca 336}
  \city{Montpellier} 
  \country{France}
}
\email{seriai@lirmm.fr}

\renewcommand{\shortauthors}{A. Shatnawi, H. Shatnawi, M. A. Saied, Z. Al Shara, H. Sahraoui, A. Seriai}

\renewcommand{\shorttitle}{Identifying Components from Object-Oriented APIs Based on Dynamic Analysis}

\begin{abstract}
The reuse at the component level is generally more effective than the one at the object-oriented class level. This is due to the granularity level where components expose their functionalities at an abstract level compared to the fine-grained object-oriented classes. Moreover, components clearly define their dependencies through their provided and required interfaces in an explicit way that facilitates the understanding of how to reuse these components. Therefore, several component identification approaches have been proposed to identify components based on the analysis object-oriented software applications.
Nevertheless, most of the existing component identification approaches did not consider co-usage dependencies between API classes to identify classes/methods that can be reused to implement a specific scenario.
In this paper, we propose an approach to identify reusable software components in object-oriented APIs, based on the interactions between client applications and the targeted API.
As we are dealing with actual clients using the API, dynamic analysis allows to better capture the instances of API usage.
Approaches using static analysis are usually limited by the difficulty of handling dynamic features such as polymorphism and class loading.
We evaluate our approach by applying it to three Java APIs with eight client applications from the DaCapo benchmark. DaCapo provides a set of pre-defined usage scenarios. The results show that our component identification approach has a very high precision. 

\end{abstract}


\maketitle

\section{Introduction}
\label{Introduction}

Decades of research have shown that developing software applications based on Application Programming Interfaces (APIs) improves software reuse by offering pre-implemented and tested functionalities \cite{shatnawi2016reverse} \cite{zibran2011useful} \cite{saied2015could} \cite{moritz2013export}. 
For Object-Oriented (OO) APIs, the basic unit is a class, which encapsulates its functionalities and specifies the public interface for using them \cite{moritz2013export}. Reusing and understanding large OO APIs, e.g., JDK and .NET framework APIs, are complex tasks due to large numbers of included classes and methods \cite{shatnawi2016reverse} \cite{moritz2013export} \cite{saied2016cooperative} \cite{robillard2011field}. Considering JDK 1.8.0 as an example, we have approximately 4240 classes used to provide its functionalities to software applications \cite{liguori2014java}. Thus, it is a challenge to understand this large number of classes to identify needed functionalities \cite{uddin2012temporal}.

On the other hand, API classes/methods are reused based on reuse scenarios represented by the combinations of API classes or methods that offer the required  functionalities to software applications  \cite{shatnawi2016reverse} \cite{moritz2013export}.
However, different applications rely on different scenarios, depending on their needs of API functionalities \cite{uddin2012temporal}. This leads to the possibility to have a large number of scenarios, corresponding to different combinations of API classes/methods~\cite{uddin2012temporal}. To help support understanding these reuse scenarios, several research approaches have been proposed to abstract high level views of these scenarios in terms of frequent co-usage patterns between API classes/methods. The frequent co-usage patterns are identified based on the analysis of how software applications (called client applications) have (re)used API classes/methods \cite{shatnawi2016reverse} \cite{saied2015mining} \cite{uddin2012temporal} \cite{montandon2013documenting} \cite{saied2015observational} \cite{Monperrus:2010:ECOOP} \cite{salman2017identification}.
These abstracted co-usage patterns of API classes/methods are used to support software engineers to perform several engineering tasks; API reengineering \cite{shatnawi2016reverse}, API documenting \cite{montandon2013documenting}, API understanding \cite{saied2015observational}, API policy enforcing~\cite{riganelli2017policy}~\cite{riganelli2017verifying}, etc.

Identifying components in OO APIs is another alternative to improve APIs' documentation by providing an abstract high-level view of the provided funcltionalities. In addition to the documentation motivation, the identified components can also help migrating OO APIs into component-based ones if desired. Such a reegineering process does not only support feeding component-based repositories, but it also allows one to get the benefits of the component-based software engineering by developing large-scale software applications using selected components and integrating them together based on flexible architectures \cite{shatnawi2013mining}. It has been admitted that software components are more reusable and understandable software modules than object-oriented classes. This admission is based on the granularity level where components are coarse-grained modules compared to the fine-grained object-oriented classes \cite{adjoyan2014service}. Furthermore, components clearly define their dependencies through their provided and required interfaces in an explicit way that facilitates the understanding of how to reuse these components.
These are the motivations of several component identification approaches that have been proposed to identify components based on the analysis of source code of OO software \cite{seriai2014enactment} \cite{shatnawi2013mining} \cite{allier2011object}. 

Existing component identification approaches are designed to identify components from OO source code of software applications. They only rely on OO dependencies (e.g., method invocations, sharing types, etc.) to identify dependencies between classes. However, in the context of software OO APIs, co-usage dependencies between API classes should also be considered to identify classes/methods that can be reused to implement a specific scenario. This requires the analysis of source code of APIs and their client applications. Therefore, approaches designed for OO applications cannot be applied directly to OO APIs.
To the best of our knowledge, the only approach proposed to identify components from OO APIs is one described in \cite{shatnawi2016reverse} and \cite{shatnawi2014mining}. This approach is based on the static analysis of the source code of OO APIs and their client applications. As we are dealing with actual clients using the OO API, dynamic analysis allows to better capture the instances of API usage. Approaches using static analysis are usually limited by the difficulty of handling dynamic features such as polymorphism and class loading. 

In this paper, we propose an approach to identify reusable components in OO APIs. Our approach is based on the dynamic analysis of interactions between client applications and the targeted OO API. Thus, we generate the execution traces for the different usage scenarios implemented in client applications. These execution traces realize dependencies between API classes through their method invocations. 
We assume that components are identified in terms of API classes that are frequently reused together by client applications. Therefore, we consider that groups of methods frequently appearing together in execution traces form provided interfaces of components, and the owner API classes of these methods constitute the structure of that candidate component.
To evaluate our approach, we applied it to three Java APIs with their clients from the DaCapo benchmark. The results show that the precision of our component identification approach is 98\%.

The rest of this paper is organized as follows. 
Section \ref{sec2:ApproachFramework} presents the framework behind our approach. 
Section \ref{sec3:IdentifyingExecutionTraces} describes the process of identifying execution traces related to usage scenarios of client  applications of APIs. 
In Section \ref{sec4:qualityFunction}, we identify graph representations of APIs methods based on the relationships appeared in execution scenarios using a proposed quality function.
Section \ref{sec5:identifingGroups} shows how classes composing components are identified based on a graph-based clustering algorithm. 
Evaluation results are presented in Section \ref{sec5:Experimentation}. 
Related works are discussed in Section \ref{sec6:RelatedWork}.
 Conclusion and future work are presented in Section \ref{sec8:ConclusionFutureWork}.

\section{The Proposed Approach Framework}
\label{sec2:ApproachFramework}

In this section, we provide an overview, summarize the principles and propose the process of the proposed approach.

\subsection{Approach Overview} 
Our approach identifies reusable components based on the dynamic analysis of interactions between client applications and the targeted API.
We define a software component as a collection of classes that participate to implement one or more functionalities for client applications of an API. The provided interfaces, of this component, are API methods that have been invoked together frequently by client applications. Conversely, the required interfaces are API methods that have been invoked by the component classes and belong to other API components' classes.

Classes composing a component are identified based on their invoked methods (provided interfaces). 
To determine which methods are invoked jointly, we analyze execution traces of client applications using the API. As for dynamic analysis, we need representative usage scenarios, we rely on use cases of client applications to identify such representative scenarios, and execute them to produce the traces. The execution traces highlight the dependencies between API methods. We consider methods that appear frequently together in execution traces as provided interfaces of a component. The owner classes of these methods define the structure of the component implementation.

We want to package each collection of API methods frequently used together to form a provided interface of a reusable component, without changing the internal structure of the API (i.e., we do not aim to identify architectural view of the API).
Therefore, we allow a class to be a part of more than one component as different subsets of its methods can participate with various groups of methods related to other classes to implement different functionalities.

To evaluate the quality of a candidate collection of API methods to form a component provided interface, we define quality function that analyzes dependencies between API methods based on the identified execution traces. We cluster API methods using this quality function.



\subsection{Approach Principles} 
The principles of our approach are summarized as follows.


\begin{itemize}
	\item A component is defined as a group of classes identified through its potential provided interfaces.

	\item A provided interface of a component is a set of methods used frequently together in execution traces. 

	\item A required interface of a component is a set of methods used by this component and belonging to other components' classes.

\item A class can belong to several components since different subsets of its methods can be included in different component interfaces. 

	\item Execution traces are used to identify the dependencies between API methods. Thus, they guide the component identification process.

\end{itemize}

\begin{figure}[h]
	\begin{center}
	\includegraphics[width=0.47\textwidth]{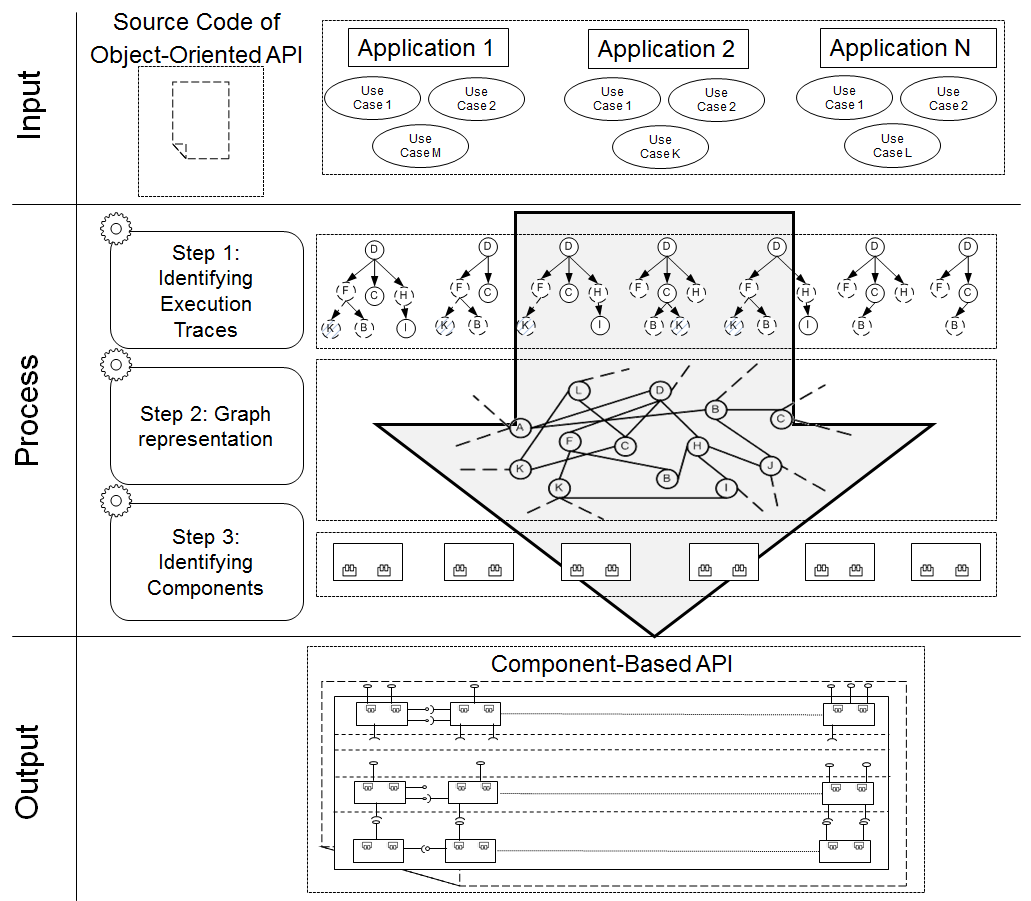}
		\caption{The process of component identification from object-oriented APIs}
		\label{fig:mining_process}
	\end{center}
\end{figure}

\subsection{Approach Process} 
To implement our approach, we propose a process, presented in Figure \ref{fig:mining_process}, based on the following three steps:

\begin{enumerate}
	\item \textbf{Identifying execution traces}:
	execution traces are identified based on usage scenarios related to client applications of APIs. Each executing trace is realized in terms of a call tree that represents the dynamic relationships between API methods corresponding to a usage scenario.
	
	\item \textbf{Building graph representations of APIs:}
	to facilitate the problem definition, we rely on graph representations to highlight dependencies between methods of an API. These dependencies are identified based on the strength of relationships between methods in the call trees. 
	 
	\item \textbf{Identifying classes composing components:}
    we apply a graph-based clustering algorithm to partition the graph into sub-graphs. Each sub-graph represents a group of methods forming a provided interface of a component, while their classes represent the component implementation.
\end{enumerate}

\section{Identifying Execution Traces}
\label{sec3:IdentifyingExecutionTraces}

In this section, we discuss how to identify execution traces related to APIs based on usage scenarios of client applications. 

An execution trace is a set of methods that represent the execution of a usage scenario of a client application of the targeted API. These methods can be represented in terms of a call tree defined as a \textit{directed tree T = $\langle V, E\rangle$}, where \textit{V} is a set of vertices referring to a set of API methods and \textit{E} is a set of edges. An edge $\langle V1, V2\rangle$ refers to that a method \textit{V1} invokes a method \textit{V2}. The root of the tree represents the starting point of the execution trace.
    
\subsection{Executing Usage Scenarios to Identify Call Trees}

We execute each usage scenario to identify the set of methods corresponding to its implementation. 
A method representing the entry of a usage scenario is considered as the call tree root. 
Then, we follow method calls to identify the other nodes. When a method invokes a method, a new node is added to the tree with an edge from the first node to the class node, and so on.

\begin{figure}
	\begin{center}
	\includegraphics[width=0.47\textwidth]{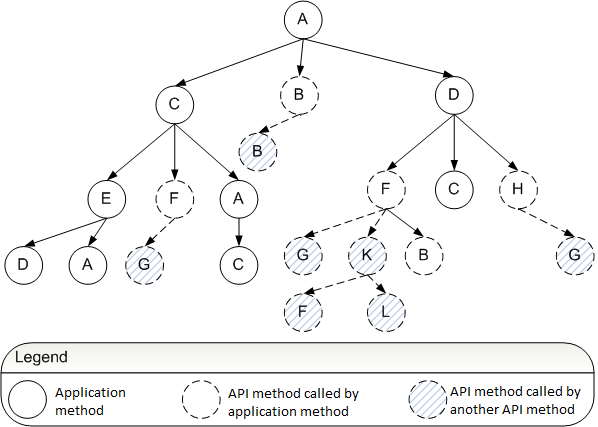}
	\caption{Example of call tree corresponding to usage scenario}
	\label{fig:api_tree}
\end{center}
\end{figure}

Methods related to the call tree's nodes can be classified into three categories: methods implemented in application's classes, method of API classes used by application's classes to access API functionalities and method of API classes that have got called by the second category's methods (from API method used by method of application's classes). 
Figure \ref{fig:api_tree} shows an example of a call tree that is composed of 4 application methods, which are \textit{A, C, D} and \textit{E}, and 6 API methods, which are \textit{B, F, G, H, L} and \textit{K}. 
The \textit{B} and \textit{F} methods  have been invoked from an application class, while \textit{G} and \textit{K}  methods  have been called from an API method.


\subsection{Removing Application's Methods from Call Trees}

\begin{figure}
	\begin{center}
	\includegraphics[width=0.47\textwidth]{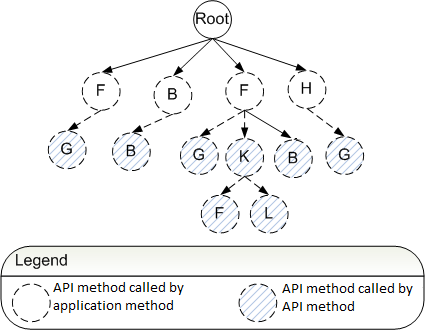}
		\caption{Pruned call tree of one in Figure \ref{fig:api_tree}}
	\label{fig:norm_api_tree}
\end{center}
\end{figure}

Since our goal is to analyze interdependencies between API methods, we prune the identified call trees by removing methods related to application classes.
For each call tree, we perform a \textit{Breadth First Search} algorithm starting from the root node. If a child is an API method, then we leave it. Otherwise, we remove the node. The children of the removed node are attached to the parent of the removed node. This process is continued until verifying all tree levels. For example, Figure \ref{fig:norm_api_tree} shows the pruned tree of the call tree presented in Figure \ref{fig:api_tree}. As it is noticed, the tree size is significantly reduced which decreases the complexity of analyzing call trees. For instance, it is reduced from 21 to 13 nodes.


\section{Building Graph Representations of APIs}
\label{sec4:qualityFunction}

We represent an API in terms of an \textit{undirected weighted graph G = $\langle V, E\rangle$}, where \textit{V} is the set of API methods and \textit{E} is the set of edges between the vertices \textit{V}. An edge links the vertices \textit{(u,v)} if they appeared together in call trees.
The weight associated to each edge is based on the strength of  relationship of its vertices in call trees. 
We distinguish three attributes that should be analyzed in call trees to measure the strength of these relationships. These are call frequency, call distance and call weight.

\subsection{Call Frequency} 

\begin{figure}
	\begin{center}
   		\includegraphics[width=0.50\textwidth]{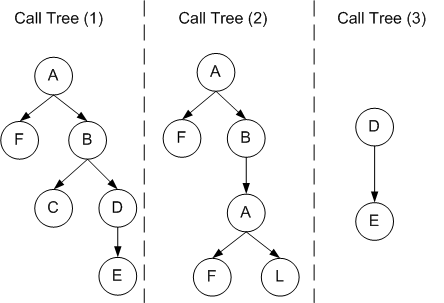}
			\caption{An example of three call trees}
		\label{fig:call_trees_example}
\end{center}
\end{figure}

This attribute refers to the relationship between numbers of co-occurrence of a group of methods in call trees and their cohesive. Methods frequently appeared in call trees are likely participating to provide related functionalities. For instance, in the call trees presented in Figure \ref{fig:call_trees_example}, the call frequency of \textit{A} and \textit{B} is 2 times, and it is 1 for \textit{A} and \textit{L}. Thus, \textit{A} and \textit{B} are likely cohesive than \textit{A} and \textit{L}.

The call trees are identified from different applications developed by different development teams. 
Thus, we propose to measure the call frequency based on two cases related to methods frequently used together in call trees of the same applications by the same team or different applications of different teams. We consider that methods used by different development teams will provide a global view of their reuse frequency (i.e., global frequency) compared to methods used by the same development team (i.e., local frequency).

Local frequency measures how many methods are used together in the same applications (the same developers). It is measured based on the average number of appearances of a group of methods in one application. We calculate the ratio between the number of call trees containing the methods to the total number of call trees in each application (c.f. Equation \ref{eq:Lfreq}). 

Global frequency measures how many methods are used together in different applications. Global frequency is calculated based on the average number of applications that contain the methods to the total number of applications (c.f. Equation \ref{eq:Gfreq}).
The call frequency of a group of methods is the mean value of local and global frequencies of all pairs of methods. 

Equation \ref{eq:callFreq} measures the call frequency for a set of methods \textit{E}, where \textit{Co-occur(c,v)} returns 1 if the methods \textit{c} and \textit{v} exist in the call tree \textit{t}, otherwise it returns 0. \textit{T(a)} refers to the set of call trees of the application \textit{a} in the set of applications \textit{apps}.

\begin{dmath}
\label{eq:callFreq}
CallFreq(E)= \frac{\sum_{c,v\epsilon{E}, c\neq{v}} \frac{LFreq(c,v)+GFreq(c,v)}{2}}{(|E|*(|E|-1))/2}
\end{dmath}

\begin{dmath}
\label{eq:Lfreq}
LFreq(c,v)= \frac{\sum_{a\epsilon{apps}} \frac{\sum_{t\epsilon{T(a)}} Co-occur(c,v)}{|T(a)|}}{|apps|}
\end{dmath}

\begin{dmath}
\label{eq:Gfreq}
GFreq(c,v)= \frac{Number Of Apps Containing(c,v)}{|apps|}
\end{dmath}

\subsection{Call Distance}

This attribute is related to the \textit{distance} between methods in call trees, which refers to the strength of their interactions. As much as methods are closer in call trees they have high probabilities to provide same functionalities.
For example, in the first call tree in Figure \ref{fig:call_trees_example}, the distance between \textit{A} and \textit{B} is 1, while it is 3 between \textit{A} and \textit{E}. Thus, \textit{A} and \textit{B} have higher probability to be cohesive than \textit{A} and \textit{E}.

We define Equation \ref{eq:CallDist} to measure the call distance of a group of methods \textit{E}. For each pair of methods, it calculates the mean distance of this pair in all call trees using Equation \ref{eq:Distance}. Then, the call distance is the average value of the mean values of all pairs.

\begin{dmath}
\label{eq:CallDist}
CallDist(E)= \frac{\sum_{c,v\epsilon{E}, c\neq{v}}Distance(c,v)} {(|E|*(|E|-1))/2}
\end{dmath}

\begin{dmath}
\label{eq:Distance}
Distance(c,v)= \frac{\sum_{a\epsilon{apps}} \frac{\sum_{t\epsilon{T(a)}} dis(c,v,t)}{|T(a)|}}{|apps|}
\end{dmath}

\begin{dmath}
	\label{eq:dis}
	dis(c,v,t)= 1-(\frac{avgDistance(c,v,t)}{2*getTreeDepth(t)})
\end{dmath}

Where \textit{avgDistance(c,v,t)} finds the average distance between the methods \textit{c} and \textit{v} in a call tree \textit{t}. 
It returns  \textit{2*getTreeDepth(t)} if the methods \textit{c} and \textit{v} do not appear in a call tree.

\subsection{Call Weight}

This attribute is related to the weight of the co-occurrence of a group of methods in call trees. Methods that are co-located in the same call trees have different weight of co-occurrences. This affects to the degree of their interaction, and consequently their cohesive. The weight depends on the number of calls in the call trees. For example, in Figure \ref{fig:call_trees_example}, \textit{D} and \textit{E} appear together in call tree (1) and call tree (3), but with different weights. This means that their appearance has different impact to their cohesive. In fact, it is clear that their appearance in call tree (3) has a higher impact than in call tree (1).

We measure the call weight based on the number of edges in call trees. For a group of methods, the call weight is the average call weight of all pairs in this group. Since the call weight of a pair of methods in such a call tree is the fraction between the number of direct call between these methods and the total number of edges in the call tree. For our example (D and E), it is 20\% in call tree (1) and 100\% in call tree (3), thus it is 60\% in average. Equation \ref{eq:callWght} measures the call weight of a collection of methods \textit{E}:

\begin{dmath}
\label{eq:callWght}
CallWeight(E)= \frac{\sum_{c,v\epsilon{E}, c\neq{v}}Weight(c,v)}{(|E|*(|E|-1))/2}
\end{dmath}

\begin{dmath}
\label{eq:wight}
Weight(c,v)= \frac{\sum_{a\epsilon{apps}}\sum_{t\epsilon{T(a)}} Wei(c,v,t)}{|apps|}
\end{dmath}

Where \textit{Wei(c,v,t)} is the wight between the methods \textit{c} and \textit{v} in the call tree \textit{t}. It is calculated using Equation \ref{eq:wi}.

\begin{dmath}
\label{eq:wi}
Wei(c,v,t)= \frac{NumberOfDirectCall(c,v,t)}{TotalNumberOfCall(t)}
\end{dmath}

For evaluating the quality of a group of methods \textit{E}, we define Equation \ref{eq:qualityFunction} as a linear combination of the three functions evaluating each attribute.

\begin{dmath}
\label{eq:qualityFunction}
  Q(E)=\frac{1}{\sum_{i}\lambda_i}\cdot(\lambda_1\cdot CallFreq(E) + \lambda_2\cdot CallDist(E) + \lambda_3\cdot CallWght(E))
\end{dmath}

Where $\lambda_1$, $\lambda_2$ and $\lambda_3$ are weight values, situated in [0-1]. These are used by the API expert to weight each attribute as needed.


\section{Identifying Components Through their Provided Interfaces}
\label{sec5:identifingGroups}

As we mentioned previously, methods composing provided interfaces of components are grouped based on a graph-based clustering algorithm. We select a  clustering algorithm that allows overlapping between clusters. The idea behind that is to allow a method to appear in different clusters (component provided interfaces) as when it is used frequently with different groups of method.

\subsection{Identifying Overlapping Clustering of Methods}
The identification of methods composing component interfaces is based on the identification of a set of subgraphs, where they have accepted quality values. 
We define the component provided interface as a subgraph $G^*$ = $\langle V^*$, $E^* \rangle$ where the average weight of edges in $E^*$ maximizes the quality function, the set of methods corresponding to the vertices $V^*$ represents the component provided interfaces.
The identification of an optimal set of subgraphs needs to identify all possible candidate subgraphs that can be extracted from the graph. Then, the selection of subgraphs is made by choosing ones that maximize the quality function. Nevertheless, this is considered as \textit{NP-Complete} problem since the time complexity is exponential. Thus, we propose to use a heuristic clustering algorithm to find a near optimal set of subgraphs. We use the \textit{OClustR} \cite{perez2013oclustr} clustering algorithm. We select this algorithm as it is a graph-based clustering algorithm that is able to identify overlapping clusters. In addition, there is no need to identify the number of the needed clusters as we do not have any idea about the number of components that should be identified.

Based on \textit{OClustR}, the problem is to identify a set of subgraphs \textit{W = $\{G_{1}^*, G_{2}^*, ..., G_{k}^*\}$}, where the graph \textit{G} is covered by \textit{W} (i.e. \textit{$ \cup_{i=1}^k V_{i}^*$ = V}, such that \textit{$G_{i}^*$} covers a vertex \textit{v} if \textit{v} belongs to \textit{$V_{i}^*$}). Finding the minimum number of subgraphs, i.e. the set \textit{W}, is known as a vertex cover problem. A vertex cover is a smallest subset of vertices \textit{V^} $\subseteq$ \textit{V}, such that each vertex \textit{v} $\in$ \textit{V} is either a part of \textit{V^} or directly connected, i.e. an adjacent vertex, to another vertex \textit{u}  $\in$ \textit{V^}. Nevertheless, a vertex cover is classified as \textit{NP-Complete} problem [Ref]. Thus, \textit{OClustR} proposed an approximation algorithm that aims at mining a near optimal set of vertices \textit{V^}, such that graphs corresponding to \textit{V^} cover \textit{G} and maximizing the quality of the corresponding component as well. This algorithm identifies the subgraphs based on two steps. The first step aims at mining the initial subgraphs (clusters). The second step goals at refining the initial subgraphs by minimizing the number of clusters, since they may include useless subgraphs, and the overlapping between them. 

\subsubsection{Mining initial clusters}
This step aims at identifying the set of initial subgraphs \textit{W}, such that each subgraph \textit{$G_{i}^*$} is a \textit{weighted star subgraph, (ws-graph)} for short, in G. A \textit{ws-graph} \textit{$G_{i}^* = \langle V^*,E^*\rangle$} have a vertex \textit{c} $\in$ \textit{$V^*$}, \textit{c} is called the \textit{center} of \textit{$G_{i}^*$}, where there is an edge connecting \textit{c} with the other vertices in \textit{$V^*$}, the other vertices are called \textit{satellites}. Here, the problem is to identify a set of \textit{center} vertices \textit{$Z = \{c_{1}, c_{2}, ..., c_{k}\}$}, where each vertex \textit{$c_{i} \in Z$} is the center of the subgraph \textit{$G_{i}^* \in W$}. In order to identify the set \textit{Z}, all vertices should be investigated. The investigation is based on  an iterative process. At each iteration, one vertex \textit{v} is added to \textit{Z}. This is continued until Z reaches the condition of covering \textit{G}. The selection of a vertex \textit{v} to be added to \textit{Z} is based on an evaluation criterion. This criterion depends on two factors. The first one aims at maximizing the cover of \textit{G} at each iteration and controlling the overlapping between the \textit{ws-graphs}. This factor is called \textit{Relative Density}, \textit{(RD)} for short. The second factor goals at keeping the quality of the component corresponding to the \textit{ws-graph} related to a vertex \textit{v}, since a vertex that provides a high coverage of G could produce a low quality component corresponding to our quality fitness function. This factor is called \textit{Relative Compactness}, \textit{(RC)} for short.

\textit{RD} is related to the number of \textit{satellite} vertices that could be covered by \textit{ws-graph} corresponding to a \textit{center} vertex \textit{v}. Since vertices having higher degree, i.e. vertices having more \textit{satellites}, produce \textit{ws-graphs} that contribute more to cover \textit{G}, the selection of a vertex should take the vertex degree into account. In some cases, this way is not sufficient since some of the \textit{satellite} vertices maybe already covered by another \textit{center} vertex \textit{r} that is added to \textit{Z} in a previous iteration. Thus, we need to take the number of \textit{satellite} vertices that will be covered by selecting a vertex \textit{v} in the current iteration, i.e. we exclude \textit{satellite} vertices already covered by another \textit{center} vertices. We measure \textit{RD} of a vertex \textit{$v \in V$} based on Equation \ref{algo.RD}. The higher value of \textit{RD} the better vertex to be selected, where its value is situated in [0-1].

\begin{dmath}
\label{algo.RD}
  RD(v)=\frac{Number Of Uncover Satellite Vertices}{Total Number Of Satellite Vertices}
\end{dmath}

\textit{RC} is related to the quality of a \textit{ws-graph} corresponding to \textit{v} compared to its \textit{satellite} vertices. This means that we investigate if a vertex \textit{v} is the best vertex that maximizing the quality among its \textit{satellite} vertices. Therefore, we measure \textit{RC} based on the ratio between the number of \textit{satellite} vertices that produces \textit{ws-graphs} having lower quality values than the \textit{ws-graph} corresponding to \textit{v} and the total number of \textit{satellite} vertices. Where the quality of a \textit{ws-graph} is the average weight values between all pairs of vertices included by the \textit{ws-graph}. We calculate \textit{RC} based on Equation \ref{alg.RC}, where \textit{SatelliteCompactness} is the number of \textit{satellites s} such that the quality of \textit{ws-graph} of \textit{s} is grater than the quality of the \textit{ws-graphs} of \textit{v}. The higher value of \textit{RC} the better vertex to be selected, where its value  is situated in [0-1].

\begin{dmath}
\label{alg.RC}
  RC(v)=\frac{SatelliteCompactness}{Total Number Of Satellite Vertices}
\end{dmath}

Based on these factors, we firstly sort the vertices in a decreasing order based on their relative quality; the average of their \textit{RD} and \textit{RC}, denoted by RQ. Then, these vertices are iteratively added to the set \textit{Z} with respect to one of these conditions: (1) a vertex is not covered and (2) it is covered but at lease there is one of its \textit{satellites} that is not covered. This continues until covering graph \textit{G}.  

\begin{algorithm}
\SetAlgoLined
\label{algo:InitialClusters}
\SetKwFunction{match}{match}
 \KwIn{undirected weighted graph \textit{G = $\langle V, E\rangle$}}
 \KwOut{A Set of \textit{center} vertices \textit{$Z$}}
 \textit{L} $\longleftarrow$ \textit{G(V)}\;
 Sort \textit{L} in a decreasing order based on \textit{RQ}\;
 \For{each \textit{v} $\in$ \textit{L}}
 {
     \If{\textit{v}.uncovered() || \textit{v}.hasUncoveredsatellites()}{
     	\textit{Z $\longleftarrow$ Z $\cup$ $\{v\}$}\;
     }
 }
\KwRet{\textit{Z}\;}
\caption{Identifying initial clusters}
\end{algorithm}

\subsubsection{Refining the initial clusters}
This step aims at enhancing the initial clusters, identified in the previous step. The goal is at reducing the number of resulted clusters as well as the overlapping between them. The process of identifying the set \textit{Z} is a greedy one since it selects the vertex having the highest quality at each iteration. Thus, this may lead to the situation of adding a useless vertex \textit{u} to \textit{Z}. The identification of a useless subgraph \textit{$G_{u}^*$} is based on how much its \textit{satellites} are shared with other \textit{$G^*$s}.
The worst case is that \textit{u} $\in$ \textit{Z}, such that \textit{G} is still covered by the \textit{$G^*$s} corresponding to \textit{$Z - \{u\}$}, i.e. \textit{u} and its \textit{satellites} are covered by other \textit{$G^*s$}. The average case is that \textit{u} $\in$ \textit{Z}, such that \textit{$G_{u}^*$} shares most of its \textit{satellites} with the other \textit{$G^*s$} corresponding to \textit{$Z - \{u\}$}. In these case, the  \textit{$G_{u}^*$} is considered as a useless subgraph and \textit{u} should be deleted from \textit{Z}. 

According to that, \textit{$G_{u}^*$} is considered as a useless if it meets two conditions. The first one is that \textit{u} is a \textit{satellite} vertex of at lease another \textit{$G_{v}^*$}. The second condition is that \textit{$V_{u}^*$} shares more than half of its \textit{satellite} vertices with another \textit{$G^*s$}. Once a useless \textit{$G_{u}^*$} is identified, \textit{u} is removed from \textit{Z}, and the non-shared \textit{satellites} are distributed to another \textit{$G^*$}. We select the \textit{$G^*$} that covers \textit{u} as it covers the center of \textit{$G_{u}^*$}. In case that there is more than one \textit{$G^*$} covering \textit{u}, we select \textit{$G^*$} having the grater number of \text{satilletes}. By doing this, we allow producing clusters having many vertices.

Algorithmically, we refine the initial clusters \textit{$G^*$} as follows. It firstly sorts the vertices of \textit{Z} in a descending order based on their degrees and sign all of them as unvisited. Then, starting from the vertex having the highest degree, each vertex \text{v} is analyzed. The analysis consists of removing from \textit{Z} any vertex \textit{$u \in Z$} such that \textit{u} is a \textit{satellite} of \textit{$G_{v}^*$} (\textit{$u \in V_{v}^*$}) and \textit{$G_{u}^*$} is considered as useless \textit{$G^*$}. Then, all \textit{satellite} verices of \textit{$G_{u}^*$} is added to \textit{$G_{v}^*$} since \textit{$G_{v}^*$} is the one having the grater number of \text{satilletes} among \textit{u} adjacent vertices. Once all \textit{satellite} verices in \textit{$G_{v}^*$} are analyzed, \textit{v} is signed as visited and \textit{$G_{v}^*$} is considered as a final cluster.

\begin{algorithm}
\SetAlgoLined
\label{algo:RefiningClusters}
\SetKwFunction{match}{match}
 \KwIn{A Set of \textit{center} vertices \textit{$Z$}}
 \KwOut{A Set of Clusters \textit{$C$}}
 \textit{L} $\longleftarrow$ \textit{G(V)}\;
 Sort \textit{Z} in a decreasing order based on their degree\;
 \For{each \textit{v} $\in$ \textit{Z}}
 {
 	\textit{v}.isVisited(\textit{false})
 }
 \For{each \textit{v} $\in$ \textit{Z}}
 {
 	\For{each \textit{u} $\in$ \textit{satellites} of \textit{$G_{v}^*$}}
 	{
 		\If{\textit{u} $\in$ \textit{Z} $\&\&$ \textit{u}.isNotVisited()}
 		{
 			\If{\textit{$G_{u}^*$} is useless}
 			{
 				\textit{$G_{v}^* \longleftarrow G_{v}^* \cup G_{u}^*$}\;
 				\textit{Z $\longleftarrow$ Z $-$ $\{u\}$}\;
 			}\Else
 			{
 				\textit{u}.isVisited(\textit{true})
 			}
 		}
 	}
 	\textit{C $\longleftarrow$ C $\cup$ $\{V_{v}^*\}$}\;
}
\KwRet{\textit{C}\;}
\caption{Refining initial clusters}
\end{algorithm}

\begin{table}[]
	\centering
	\caption{Data set description}
    \resizebox{0.5\textwidth}{!}{
	\begin{tabular}{|c|c|c|c|}
		\hline
		API name &  Size (classes)  &\# of app. clients &  App. client names  \\ \hline
		ACL &          23     &       2             & Tomcat, Fop  \\ \hline
		ASM      &         99	    &     2        &  Jython, Pmb  			 \\ \hline
		XML      &         302      &     4        &  Xalan, Pmd, Fop, Batik  \\ \hline
	\end{tabular}
    }
	\label{tb:dataSet}
    
\end{table}

\begin{table*}
	\centering
	\caption{The results of call trees identification}
	\resizebox{\textwidth}{!}{
	\begin{tabular}{|l|c|c|c|c|c|c|}
		\hline
		\multicolumn{1}{|c|}{Call tree} & Total no. of nodes & No. of API unique methods    & Tree height & Min method repetition & Max method repetition 	& Average method repetition\\ \hline
		ACL1                            &   1983448          &    40					    &  12         &      1              & 971877          		& 49586.2						\\ \hline
		ACL2                            &   467328           &   81			                &   13        &      1              & 233231           		& 5769.48       				 \\ \hline
		XML1                            & 79844              & 57                  			& 4           & 1                   & 11040             	& 1400.77						\\ \hline
		XML2                            & 116013             & 79                 			& 7           & 1                   & 25392             	& 1468.52						 \\ \hline
		XML3                            & 245                & 25                			& 14          & 1                   & 20                	&	9.8							\\ \hline
		XML4                            & 1694602            & 84                 			& 5           & 1                   & 769600             	&	20173.83					\\ \hline
		ASM1                            & 3366089            & 106                 			& 10          & 1                   & 589875             	&	31755.56					\\ \hline
		ASM2                            & 527021             & 28                  			& 27          & 1                   & 232285             	&	18822.18					\\ \hline		
	\end{tabular}
	\label{tb:allcalltrees}
}
\end{table*}

\section{Evaluation Results}
\label{sec5:Experimentation}

\subsection{Description of APIs and their Client Applications}
We evaluate our approach using real usage scenarios from DaCapo Benchmarks \cite{blackburn2006dacapo}. We select DaCapo because it is composed of a set of open-source real Java applications coupled with collections of \textit{pre-defined usage scenarios}.

To show the applicability of our approach in several context, we select from DaCapo three widely used Java APIs of three different size (i.e., 23, 99 and 302 classes) and from three different domains. 
The first one is the Apache Commons Logging (ACL) API\footnote{Available at: https://commons.apache.org/proper/commons-logging/guide.html} that contains 23 classes, and offers a log interfaces and a middleware/tooling developer with a simple logging abstraction. 
The second one is the ASM API \footnote{Available at: http://www-etud.iro.umontreal.ca/~saiedmoh/asm-3.3.1/doc/javadoc/user/index.html} 
consists of 99 classes. It provides functionalities related to Java  bytecode manipulation and analysis.
The third API is the XML API\footnote{Available at: https://xerces.apache.org/xerces2-j/javadocs/api/index.html} that is composed of 302 classes. It offers functionalities for processing XML files.

DaCapo only provides a limited number of usage scenarios of software applications developed based on these three APIs. We are able to identify only eight client applications that use methods of classes of the APIs.  
These applications are as follows. 

\begin{itemize}
\item \textbf{Tomcat} implements J2EE technologies like Servlet and JavaServer Pages. 

\item \textbf{Fop} is an output-independent print formatter that parses and formats XSL-FO files. 

\item \textbf{Xalan} is an XSLT processor for transforming XML documents.

\item \textbf{Batik} is a Scalable Vector Graphics (SVG) toolkit that renders a number of SVG files. Xalan, Fop and Batik use classes from the XML API. 

\item \textbf{Pmd} is a source code analyzer for Java code. It uses two APIs; XML and ASM ones. 

\item \textbf{Jython} is a python interpreter written in Java to execute and interpret Python programs. It is a client for the ASM API.

\item \textbf{Lusearch} and \textbf{Luindex} are respectively a text search tool and a text indexing for a corpus of data comprising the works of Shakespeare and the King James bible. Both applications are clients of the Licene API.
\end{itemize}

Table \ref{tb:dataSet} shows the description of the collected dataset. For each API, we provide the number of included classes, the number of applications considered as clients of this API and the names corresponding to the client applications.

\subsection{Evaluation Process}
Our evaluation process is based on four steps. First, we have discussed the results of identifying call trees related to execution traces of usage scenarios of client applications of APIs. Then, we have shown the results of identifying groups of methods that are considered as provided interfaces of components. Next,we have presented how we evaluate the resulting components based on functionalities provided by methods representing their provided interfaces. Next, we have provided a discussion about dynamic and static analysis component identification approaches. Finally, we have discussed threats to validity related to the results of our approach.

\begin{table}[h]
	\centering
	\caption{The results of component identification}
        \resizebox{0.5\textwidth}{!}{

	\begin{tabular}{|c|c|c|c|}
		\hline
		API & \begin{tabular}[c]{@{}c@{}}\# of identified \\ components\end{tabular} & \begin{tabular}[c]{@{}c@{}}Avg. provided interface \\ size (methods)\end{tabular} & \begin{tabular}[c]{@{}c@{}}Avg. component\\ size (classes)\end{tabular} \\ \hline
		ACL & 9                                                                      & 11                                                                                & 2.14                                                                    \\ \hline
		ASM & 24                                                                     & 5.08                                                                              & 2.13                                                                    \\ \hline
		XML & 22                                                                     & 6.72                                                                              & 3.25                                                                    \\ \hline
	\end{tabular}
    }
	\label{tb:componentIdentification}
\end{table}

\begin{table}[h]
	\centering
	\caption{Examples of identified components}
	\resizebox{\textwidth/2}{!}{
		\begin{tabular}{|c|l|}
			\hline
			API    & Component provided interface names\\
			\hline
			ACL    & \begin{tabular}[c]{@{}l@{}}org.apache.commons.logging.impl.SimpleLogrun,\\ org.apache.commons.logging.impl.LogFactoryImplcreateLogFromClass,\\ org.apache.commons.logging.impl.SimpleLoggetContextClassLoader,\\ org.apache.commons.logging.impl.SimpleLog,\\ org.apache.commons.logging.impl.SimpleLogclass,\\ org.apache.commons.logging.impl.SimpleLogaccess,\\ org.apache.commons.logging.impl.SimpleLoggetResourceAsStream,\\ org.apache.commons.logging.impl.LogFactoryImpldiscoverLogImplementation,\\ org.apache.commons.logging.impl.LogFactoryImplnewInstance,\\ org.apache.commons.logging.impl.SimpleLog\end{tabular}   \\                                                                                                                                                                                          
			\hline
			ASM    &     \begin{tabular}[c]{@{}l@{}}org.objectweb.asm.commons.EmptyVisitorvisitCode,\\ org.objectweb.asm.commons.EmptyVisitorvisitTableSwitchInsn,\\ org.objectweb.asm.commons.EmptyVisitorvisitIntInsn,\\ org.objectweb.asm.commons.EmptyVisitorvisit,\\ org.objectweb.asm.commons.EmptyVisitorvisitMethodInsn,\\ org.objectweb.asm.commons.EmptyVisitorvisitInsn,\\ org.objectweb.asm.commons.EmptyVisitorvisitJumpInsn,\\ org.objectweb.asm.commons.EmptyVisitorvisitMethod\end{tabular}\\
			 \hline
 			XML    & \begin{tabular}[c]{@{}l@{}}javax.xml.parsers.DocumentBuilderFactoryisIgnoringElementContentWhitespace,\\ javax.xml.parsers.DocumentBuilderFactoryisNamespaceAware,\\ javax.xml.parsers.DocumentBuilderFactory,\\ javax.xml.parsers.DocumentBuilderFactoryisCoalescing,\\ javax.xml.parsers.DocumentBuilderFactoryisExpandEntityReferences,\\ org.xml.sax.InputSourcegetByteStream,\\ javax.xml.parsers.SecuritySupportgetContextClassLoader,\\ javax.xml.parsers.DocumentBuilder,\\ javax.xml.parsers.DocumentBuilderFactorynewInstance,\\ javax.xml.parsers.DocumentBuilderparse,\\ javax.xml.parsers.SecuritySupport,\\ org.xml.sax.InputSourcegetCharacterStream,\\ javax.xml.parsers.DocumentBuilderFactoryisValidating,\\ javax.xml.parsers.DocumentBuilderFactoryisIgnoringComments,\\ org.xml.sax.InputSourcegetEncoding\end{tabular} \\
			 \hline
		\end{tabular}
	}
	\label{tb:exampleComponent}
\end{table}

\subsection{Results of Identifying Call Trees Based on Executing Usage Scenarios}

We rely on BTrace\footnote{Available at: https://kenai.com/projects/btrace} dynamic tracing tool to collect execution traces related to usage scenarios. 
BTrace supports the execution of a Java program based on its bytecode and dynamically collect the methods related to this execution. We configure BTrace to run each client application only once since we are not interested in performance analysis. 

Table \ref{tb:allcalltrees} shows the results related to the identified call trees of execution traces of usage scenarios of client applications of APIs. For each call tree, we present the tree's size in terms of the number of included nodes, the number of distinct API methods included in this call tree, the tree's height, the minimum, maximum and average method' repetition in this call tree.

The results show that the call trees have small heights compared to the large number of nodes included in the trees. The average height of call trees related to ACL, XML and ASM APIs is respectively 12.5 ((12+13)/2), 7.5 ((7+4+14+5)/4) and 18.5 ((10+27)/2), while the average trees size is respectively 472676 and 1946555 nodes.
This means that execution traces go deeply by 12.5, 7.5 and 18.5 methods in average respectively for ACL, XML and ASM APIs. 

Indeed, the root of a call tree represents an execution scenario, and its children nodes represents the set of methods directly invoked in the source code of client applications to access API's functionalities. Each sub-tree that corresponds to a root child explains internal dependencies related to method invocation between API methods to provide the invoked functionality by the corresponding root child.
For example, the height of a call tree of 18 nodes refers to the maximum number of API methods that are invoked corresponding to an API method invocation directly invoked by client applications. 

Another observation is that the call trees are widely distributed in a horizontal way, i.e., each node in a call tree has a large number of children relatively to its height. Therefore, API methods have dense interactions with each others such that each API method invokes a bunch of other API methods. In addition to that, we find that some API methods have been invoked only once compared to some other ones which have been invoked 198972.5 times in average (average max node repetition). Many groups of methods are invoked in similar frequently at the same time. These are interested to be in a same provided interface, even if they belong to different classes.


\subsection{Results of Component Identification}
The results of our clustering algorithm are shown in Table \ref{tb:componentIdentification}. For each API, we present the number of identified components, the average size of provided interfaces in terms of methods  and the average component size in terms of included classes.

Respectively for ACL, ASM and XML APIs, the results show that the average number of methods that need to be used together to access an API functionality is 9, 24 and 22 methods. These methods offer functionalities of 2.14, 2.13 and 3.25 owner classes in average.

For each API, Table \ref{tb:exampleComponent} shows an example of a group of object-oriented API classes representing the implementation of identified component provided interfaces. To enable their reuse for software developers, these groups need to be transformed to confirm an exiting component model. It worths to mention some works that can be used by the user of our approach to transform these object-oriented implementation of identified components to be confirmed to component models. Alshara et al. \cite{alshara2016materializing} \cite{alshara2015migrating} provided initial approaches for transforming object-oriented component implementation to be confirmed to several component models (e.g., OSGi \cite{alliance2003osgi} and Fractal \cite{bruneton2006fractal}).

\subsection{Results of Evaluating Identified Components}
The identified components are evaluated based on the functionalities offered by their provided interfaces. For each component, we evaluate how much of the methods composing its provided interface are related to provide the same API functionalities. The quality of each provided interface is calculated based on the ratio between the number of related methods to the total number of methods composing this interface (c.f. Equation \ref{eq:evaluation}). 
We rely on the API documentations to identify each method's functionality. Then, among a group of methods composing the provided interface, we select ones that have related functionalities  to at least one method in this group. 

For a pair of methods, we consider them as functional-related if one of these two cases is applied. The first one refers to identify direct indication(s) in the API documentations stated the correlation. The second one is based on human experts where their experiences allow them to decide if the two methods is functional-related.
The number of selected methods represents the numerator in Equation \ref{eq:evaluation}. The authors perform the evaluation themselves. To avoid bias, each component was evaluated by at least three authors. The average represents the final evaluation result of this component.

\begin{dmath}
	\label{eq:evaluation}
	Evaluation(Component)=\frac{No Of Related Methods}{Total No. Of Methods}
\end{dmath}

\begin{figure}[h]
	\begin{center}
		\includegraphics[scale=0.44]{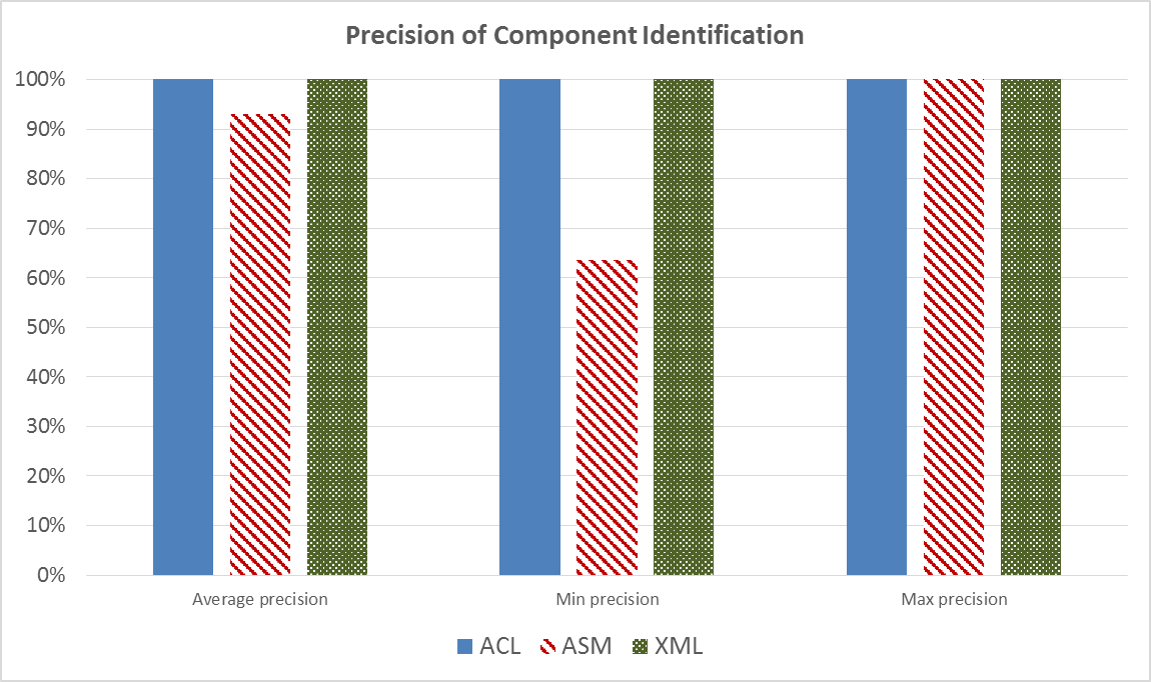}
		\caption{The precision of component identification}
		\label{fig:precision}
	\end{center}
\end{figure}

The precision of our approach is the average of the resulting values by applying Equation \ref{eq:evaluation} to components of an API. Figure \ref{fig:precision} shows the precision. The results show that our approach is able to correctly identify components where their provided interfaces are composed of functional-related API methods with 98\% precision ((100\%+93\%+100\%)/3). 
During our evaluation, we note that a component can be composed of one class and a subset of its methods forms the provided interface of this component. The same class forms another component with different subset of methods as provided interface. We consider these as correctly identified components that offer (sub)functionalities related to their provided interfaces. Each component can be (re)used as stand-alone or composed with other ones to generate higher level components (composite components).

Moreover, we observe that the same set of classes can form different components using different subsets of their methods as provided interfaces, i.e., component having the same classes but different provided interfaces. By checking the functionality of each method in these components, we find that they are related to same functionalities, but in different contexts.
As it is noticed, some classes are parts of several components since different subset of their methods are part of these components. We look to the position of these classes as correct when the corresponding methods are harmonic with other methods in a given component. In XML API, we identify a component that its provided interface consists of 14 methods belonging to 11 different classes. After checking the functionality offered by these methods, we find that methods of 7 classes are related to the same functionality, while the other 4 classes are not related to each other since they are helper classes.



\subsection{Dynamic vs Static Analysis}
The dynamic analysis approach is very expensive (run-time startup, collect usage scenarios ...), if we compare it to the static analysis which can only rely on source code files. However, dynamic analysis is more accurate as it contributes to provide real dependencies that cannot be recovered only by static analysis. Such dependencies are those related to late-dynamic binding, method overriding polymorphism and implicit dependencies in configuration files. These dependencies can exist extensively or rarely depending on the application context and framework.

On the other hand, the dynamic analysis approach relies on usage scenarios to cover API functionalities. However, some API functionalities may not be covered by the usage scenarios implemented in the selected client applications (thus, they will not be covered by the identified components). We recommend to select client applications that cover the maximum number of possible usage scenarios, which can be a strong constraint in some cases. 

We compare the two approaches by applying our approach and the static approach presented in \cite{shatnawi2016reverse} and \cite{shatnawi2014mining} into the collected APIs and client applications. We find that both approaches construct different abstractions of the same functionalities. Components identified by dynamic analysis can be composed to form components resulting from static analysis. In terms of functionality-related, the two approaches provide similar results. We interpret this similarity by the fact that the studied APIs do not rely on the dependencies that are not detected based on the static  analysis approach (e.g., implicit dependencies).
In this context, we would like to recommend to rely on the dynamic analysis approach for APIs that do rely on late-dynamic binding, methods overriding polymorphism and implicit dependencies in configuration files to codify dependencies. 

\subsection{Threats to Validity}
\label{ch6:threats}
We identify two types of threats to validity concerning the proposed approach; internal validity and external validity.

\subsubsection{Threats to Internal Validity}
Two aspects concern the internal threats to validity: 
\begin{enumerate}
\item We need to rely on human expert opinions to evaluate the results. However, it is not easy to access the experts of the APIs. The authors considered themselves as experts since each author has at least more than 8 years of experience in software development. 

\item Our evaluation does not consider the recall as we are not able to identify how many API methods are actually related to a functionality and not extracted by the approach. However, the recall depends on the coverage of the usage scenarios and does not rely on our approach. Any API method used in at least one usage scenario will be a part of at least one component.
\end{enumerate}

\subsubsection{Threats to External Validity}

We identified two aspects related to the external threats to validity:

\begin{enumerate}
\item We evaluated our approach based on APIs and client applications written using \textit{Java}. The proposed approach can be generalized not only for other object-oriented languages (e.g., C++), but also for procedural ones (e.g, C, Pascal).
	The idea is that our approach does not care about the implementing programing language since its input is a set of call trees. The nodes in these call trees can refer either to methods or procedures/routines. The main difference is that identified components will be groups of procedures.
    
    \item The selection of API client applications may impact identified components as different applications use API methods following different scenarios. This may impact the reusability of identified components for new independent applications.
	However, we assume API functionalities are (re)used following similar patterns by developers. This assumption was successfully utilized and proven in several research papers \cite{shatnawi2016reverse} \cite{uddin2012temporal} \cite{montandon2013documenting} \cite{Monperrus:2010:ECOOP}. We recommend selecting as much as possible of API client applications to minimize the influence of domain-specific API usages.
\end{enumerate}

\section{Related Work}
\label{sec6:RelatedWork}
To the best of our knowledge, there is only one approach proposed to identify components from object-oriented APIs \cite{shatnawi2016reverse} \cite{shatnawi2014mining}. It identifies components as groups of API classes that are frequently used together by the client applications of an API and structurally dependent. To identify co-usage relationships between classes, Frequent-Pattern Growth algorithm was used where transactions are collections of API classes used by a client application.
However, the approach presented in \cite{shatnawi2016reverse} \cite{shatnawi2014mining} does not trace the API method calls from client applications to deepen in the API as the proposed approach does to identify relationships between API classes. Instead, it relies on a static analysis technique to analyze structural dependencies in the source code of APIs. 
Recently developed software systems contains implicit dependencies resulted form the use of late dynamic-binding, Java-reflection and container-services offered by the frameworks. These  implicit  dependencies are difficult to be detected using static analysis techniques~\cite{shatnawi2017analyzing}.


All of these existing component identification approaches that are designed to identify components from OO software applications are not be directly applied to OO APIs. The reason is that they only relied on OO dependencies (e.g., method invocations, sharing types, etc.) to identify dependencies between classes, while co-usage dependencies between API classes should also be considered to identify classes/methods that can be reused to implement a specific scenario. 

In the following three paragraphs, we discuss the related work on component identification in standalone applications using dynamic and static analysis, as well as the identification of usage patterns.

\textbf{Dynamic Analysis Component Identification from Applications.}
In \cite{allier2011object}, the authors presented an approach to identify component-based architectures from object-oriented applications. The identification is based on execution traces derived from use case scenarios. Classes frequently appearing in traces represent a candidate component. Grouping these classes is done by a clustering algorithm and a heuristic search, simulated annealing. Both techniques rely on a quality function. This quality function only considers the call frequency metric, which is insufficient without including metrics like our call weight and call distance.
In \cite{cornelissen2007dynamic}, the authors presented an approach to extract an architecture view of object-oriented software using the analysis of execution traces. This research contribution lacks details on how such a process can be automated,  and only the visualization aspect is discussed.
In \cite{dugerdil2013dynamic}, the authors relied on the execution traces to recover two artifacts: 1) information that is used to update use cases' documentations, and 2) component-based architecture by mapping the traced classes into clusters. Execution traces are identified from the business tasks of end-users, i.e., the records of what classes have been executed during a user usage. 

\textbf{Static Analysis Component Identification from Applications.}
In \cite{Boussaidi2009}, the authors used the Knowledge Discovery Meta-model to represent the software elements, at different levels of abstraction, and their structural dependencies. They used a vertical clustering algorithm to identify software components as groups of classes, and a horizontal clustering algorithm to recover the layered architecture view of the identified components.
In \cite{seriai2014deriving}, the authors proposed an approach to identify component interfaces of object-oriented components identified using reverse engineering approaches. Classes of a component are analyzed to structure the required and provided interfaces. They extract a set of methods invoked by other components. These methods are then grouped by means of formal concept analysis.
Some other approaches relied on the analysis of several software applications at the same time to identify components cross these applications \cite{shatnawi2017recovering} \cite{shatnawi2015recovering} \cite{shatnawi2013mining}. For example, in \cite{shatnawi2013mining}, components are identified as groups of classes frequently presented in different applications and structural dependent. 

\textbf{Frequent Usage Patterns of API entities.} Several approaches have been proposed to identify abstract reuse scenarios in terms of frequent usage patterns of API entities. These frequent usage patterns are not themselves direct reusable entities, but they help improving the reusability and understandability of APIs.
We classified these existing approaches following their goals, usage-order consideration, the granularity of the API entities under the study, and the algorithm used to identify usage patterns.
Generally the goal of usage-pattern identification approaches can be: (i) to provide examples to support recommendation systems \cite{Montandon:2013:WCRE} \cite{Uddin:2012:TAA}, (ii) to support the documentation of APIs at different levels of abstraction \cite{Montandon:2013:WCRE} \cite{Wang:2013:MSH} \cite{saied2016cooperative}, (iii) to predict bugs resulted from incorrect usage scenarios \cite{Monperrus:2010:ECOOP}, etc.
The patterns were identified with respect to the order in which the API elements are used in some approaches \cite{Montandon:2013:WCRE} \cite{Wang:2013:MSH}, while other approaches do not take into account such an order \cite{Monperrus:2010:ECOOP} \cite{Bruch:2006:FIS} \cite{saied2016cooperative}.
The granularity of the API elements that compose the identified patterns is at the method-level \cite{Montandon:2013:WCRE} \cite{Wang:2013:MSH} \cite{saied2016cooperative} or the class-level \cite{shatnawi2016reverse}\cite{Bruch:2006:FIS}.  In  \cite{salman2017identification}, multi-level patterns are identified.
As for the algorithms to identify the patterns, the approaches used association rules mining \cite{Bruch:2006:FIS}, frequent-pattern growth \cite{shatnawi2016reverse}, clustering algorithms \cite{Wang:2013:MSH} \cite{saied2016cooperative} or a heuristic defined by the authors \cite{Montandon:2013:WCRE} \cite{Monperrus:2010:ECOOP}. Some approaches relied on a combination of many algorithms like Principle Component Analysis and clustering algorithm \cite{Uddin:2012:TAA}. 

\section{Conclusion and Future Work}
\label{sec8:ConclusionFutureWork}

We propose an approach that aims to identify reusable software components based on the dynamic analysis of interactions between client applications and the targeted API. The approach relies on co-usage and objected-oriented dependencies to define relationships between classes of APIs. To do so, we execute usage scenarios of client applications of APIs to collect execution traces. Our approach packages groups of methods frequently appearing together in the collected traces as component provided interfaces, and their owner classes define the structure of components' implementation. These groups of methods are identified based on a graph-based clustering algorithm from the information in execution traces.

To evaluate our approach, we experimented with the DaCoPo benchmarks, focusing on three of its APIs and eight client applications. We considered pre-defined usage scenarios that are already provided by DaCoPo. The evaluation results show that the precision of our approach is up to 98\%.

We will consider three future directions to investigate

\begin{enumerate}
\item We plan to transform the object-oriented implementation of the identified component provided interfaces into service-oriented interfaces to have truly reusable interfaces.

\item We want to extend the approach evaluation by considering more APIs and client applications to generalize the approach results.

\item We will also provide a visualization framework, that developers can use to identify reusable components of an API. They will have to load a set of client application using the API of interest, run the client applications while the framework is collecting the execution traces, and produces reusable components.
\end{enumerate}

\balance
\bibliographystyle{unsrt}

\bibliographystyle{ACM-Reference-Format}
\bibliography{main}

\begin{thebibliography}{10}

\bibitem{shatnawi2016reverse}
Anas Shatnawi, Abdelhak-Djamel Seriai, Houari Sahraoui, and Zakarea Alshara.
\newblock Reverse engineering reusable software components from object-oriented
  apis.
\newblock {\em Journal of Systems and Software}, 131:442--460, 2017.

\bibitem{zibran2011useful}
Minhaz~F Zibran, Farjana~Z Eishita, and Chanchal~K Roy.
\newblock Useful, but usable? factors affecting the usability of apis.
\newblock In {\em 2011 18th Working Conference on Reverse Engineering (WCRE)},
  pages 151--155. IEEE, 2011.

\bibitem{saied2015could}
Mohamed~Aymen Saied, Hani Abdeen, Omar Benomar, and Houari Sahraoui.
\newblock Could we infer unordered api usage patterns only using the library
  source code?
\newblock In {\em Proceedings of the 2015 IEEE 23rd International Conference on
  Program Comprehension}, pages 71--81. IEEE Press, 2015.

\bibitem{moritz2013export}
Evan Moritz, Mario Linares-V{\'a}squez, Denys Poshyvanyk, Mark Grechanik,
  Collin McMillan, and Malcom Gethers.
\newblock Export: Detecting and visualizing api usages in large source code
  repositories.
\newblock In {\em Proceedings of the 28th IEEE/ACM International Conference on
  Automated Software Engineering}, pages 646--651. IEEE Press, 2013.

\bibitem{saied2016cooperative}
Mohamed~Aymen Saied and Houari Sahraoui.
\newblock A cooperative approach for combining client-based and library-based
  api usage pattern mining.
\newblock In {\em 2016 IEEE 24th International Conference on Program
  Comprehension (ICPC)}, pages 1--10. IEEE, 2016.

\bibitem{robillard2011field}
Martin~P Robillard and Robert Deline.
\newblock A field study of api learning obstacles.
\newblock {\em Empirical Software Engineering}, 16(6):703--732, 2011.

\bibitem{liguori2014java}
R~Liguori and P~Liguori.
\newblock {\em Java 8 Pocket Guide}.
\newblock " O'Reilly Media, Inc.", 2014.

\bibitem{uddin2012temporal}
Gias Uddin, Barth{\'e}l{\'e}my Dagenais, and Martin~P Robillard.
\newblock Temporal analysis of api usage concepts.
\newblock In {\em 34th International Conference on Software Engineering}, pages
  804--814. IEEE Press, 2012.

\bibitem{saied2015mining}
Mohamed~Aymen Saied, Omar Benomar, Hani Abdeen, and Houari Sahraoui.
\newblock Mining multi-level api usage patterns.
\newblock In {\em 2015 IEEE 22nd International Conference on Software Analysis,
  Evolution and Reengineering (SANER)}, pages 23--32. IEEE, 2015.

\bibitem{montandon2013documenting}
Jo{\~a}o~Eduardo Montandon, Hudson Borges, Daniel Felix, and Marco~Tulio
  Valente.
\newblock Documenting apis with examples: Lessons learned with the apiminer
  platform.
\newblock In {\em 2013 20th Working Conference on Reverse Engineering (WCRE)},
  pages 401--408. IEEE, 2013.

\bibitem{saied2015observational}
Mohamed~Aymen Saied, Houari Sahraoui, and Bruno Dufour.
\newblock An observational study on api usage constraints and their
  documentation.
\newblock In {\em 2015 IEEE 22nd International Conference on Software Analysis,
  Evolution and Reengineering (SANER)}, pages 33--42. IEEE, 2015.

\bibitem{Monperrus:2010:ECOOP}
Martin Monperrus, Marcel Bruch, and Mira Mezini.
\newblock Detecting missing method calls in object-oriented software.
\newblock In {\em European Conference on Object-Oriented Programming}, pages
  2--25. Springer, 2010.

\bibitem{salman2017identification}
Hamzeh~Eyal Salman.
\newblock Identification multi-level frequent usage patterns from apis.
\newblock {\em Journal of Systems and Software}, 130:42--56, 2017.

\bibitem{riganelli2017policy}
Oliviero Riganelli, Daniela Micucci, and Leonardo Mariani.
\newblock Policy enforcement with proactive libraries.
\newblock In {\em Proceedings of the 12th International Symposium on Software
  Engineering for Adaptive and Self-Managing Systems}, pages 182--192. IEEE
  Press, 2017.

\bibitem{riganelli2017verifying}
Oliviero Riganelli, Daniela Micucci, Leonardo Mariani, and Yli{\`e}s Falcone.
\newblock Verifying policy enforcers.
\newblock In {\em International Conference on Runtime Verification}, pages
  241--258. Springer, 2017.

\bibitem{shatnawi2013mining}
Anas Shatnawi and Abdelhak-Djamel Seriai.
\newblock Mining reusable software components from object-oriented source code
  of a set of similar software.
\newblock In {\em IEEE 14th International Conference on Information Reuse and
  Integration (IRI)}, pages 193--200. IEEE, 2013.

\bibitem{adjoyan2014service}
Seza Adjoyan, Abdelhak-Djamel Seriai, and Anas Shatnawi.
\newblock Service identification based on quality metrics object-oriented
  legacy system migration towards soa.
\newblock In {\em SEKE: Software Engineering and Knowledge Engineering}, pages
  1--6. Knowledge Systems Institute Graduate School, 2014.

\bibitem{seriai2014enactment}
Abderrahmane Seriai, Salah Sadou, and Houari~A Sahraoui.
\newblock Enactment of components extracted from an object-oriented
  application.
\newblock In {\em European Conference on Software Architecture}, pages
  234--249. Springer, 2014.

\bibitem{allier2011object}
Simon Allier, Salah Sadou, Houari Sahraoui, and R{\'e}gis Fleurquin.
\newblock From object-oriented applications to component-oriented applications
  via component-oriented architecture.
\newblock In {\em 9th Working IEEE/IFIP Conference on Software Architecture
  (WICSA)}, pages 214--223. IEEE, 2011.

\bibitem{shatnawi2014mining}
Anas Shatnawi, Abdelhak Seriai, Houari Sahraoui, and Zakarea Al-Shara.
\newblock Mining software components from object-oriented apis.
\newblock In {\em International Conference on Software Reuse}, pages 330--347.
  Springer, 2015.

\bibitem{perez2013oclustr}
Airel P{\'e}rez-Su{\'a}rez, Jos{\'e}~F Mart{\'\i}nez-Trinidad, Jes{\'u}s~A
  Carrasco-Ochoa, and Jos{\'e}~E Medina-Pagola.
\newblock Oclustr: A new graph-based algorithm for overlapping clustering.
\newblock {\em Neurocomputing}, 121:234--247, 2013.

\bibitem{blackburn2006dacapo}
Stephen~M Blackburn, Robin Garner, Chris Hoffmann, Asjad~M Khang, Kathryn~S
  McKinley, Rotem Bentzur, Amer Diwan, Daniel Feinberg, Daniel Frampton,
  Samuel~Z Guyer, et~al.
\newblock The dacapo benchmarks: Java benchmarking development and analysis.
\newblock In {\em ACM Sigplan Notices}, volume~41, pages 169--190. ACM, 2006.

\bibitem{alshara2016materializing}
Zakarea Al-Shara, Abdelhak-Djamel Seriai, Chouki Tibermacine, Hinde~Lilia
  Bouziane, Christophe Dony, and Anas Shatnawi.
\newblock Materializing architecture recovered from oo source code in
  component-based languages.
\newblock In {\em ECSA: European Conference on Software Architecture}, 2016.

\bibitem{alshara2015migrating}
Zakarea Al-Shara, Abdelhak-Djamel Seriai, Chouki Tibermacine, Hinde~Lilia
  Bouziane, Christophe Dony, and Anas Shatnawi.
\newblock Migrating large object-oriented applications into component-based
  ones.
\newblock In {\em GPCE: Generative Programming: Concepts and Experiences},
  volume~51, pages 55--64, 2015.

\bibitem{alliance2003osgi}
OSGi Alliance.
\newblock {\em Osgi service platform, release 3}.
\newblock IOS Press, Inc., 2003.

\bibitem{bruneton2006fractal}
Eric Bruneton, Thierry Coupaye, Matthieu Leclercq, Vivien Qu{\'e}ma, and
  Jean-Bernard Stefani.
\newblock The fractal component model and its support in java.
\newblock {\em Software: Practice and Experience}, 36(11-12):1257--1284, 2006.

\bibitem{shatnawi2017analyzing}
Anas Shatnawi, Hafedh Mili, Ghizlane~El Boussaidi, Anis Boubaker, Yann-Ga{\"e}l
  Gu{\'e}h{\'e}neuc, Naouel Moha, Jean Privat, and Manel Abdellatif.
\newblock Analyzing program dependencies in java ee applications.
\newblock In {\em Proceedings of the 14th International Conference on Mining
  Software Repositories}, pages 64--74. IEEE Press, 2017.

\bibitem{cornelissen2007dynamic}
Bas Cornelissen.
\newblock Dynamic analysis techniques for the reconstruction of architectural
  views.
\newblock In {\em 14th Working Conference on Reverse Engineering (WCRE)}, pages
  281--284. IEEE, 2007.

\bibitem{dugerdil2013dynamic}
Philippe Dugerdil and David Sennhauser.
\newblock Dynamic decision tree for legacy use-case recovery.
\newblock In {\em 28th Annual ACM Symposium on Applied Computing}, pages
  1284--1291. ACM, 2013.

\bibitem{Boussaidi2009}
Ghizlane El~Boussaidi, Alvine~Boaye Belle, Stephane Vaucher, and Hafedh Mili.
\newblock Reconstructing architectural views from legacy systems.
\newblock In {\em 2012 19th Working Conference on Reverse Engineering (WCRE)},
  pages 345--354. IEEE, 2012.

\bibitem{seriai2014deriving}
Abderrahmane Seriai, Salah Sadou, Houari Sahraoui, and Salma Hamza.
\newblock Deriving component interfaces after a restructuring of a legacy
  system.
\newblock In {\em 2014 IEEE/IFIP Conference on Software Architecture (WICSA)},
  pages 31--40. IEEE, 2014.

\bibitem{shatnawi2017recovering}
Anas Shatnawi, Abdelhak-Djamel Seriai, and Houari Sahraoui.
\newblock Recovering software product line architecture of a family of
  object-oriented product variants.
\newblock {\em Journal of Systems and Software}, 131:325--346, 2017.

\bibitem{shatnawi2015recovering}
Anas Shatnawi, Abdelhak Seriai, and Houari Sahraoui.
\newblock Recovering architectural variability of a family of product variants.
\newblock In {\em International Conference on Software Reuse}, pages 17--33.
  Springer, 2015.

\bibitem{Montandon:2013:WCRE}
J.E. Montandon, H.~Borges, D.~Felix, and M.T. Valente.
\newblock Documenting apis with examples: Lessons learned with the apiminer
  platform.
\newblock In {\em 20th Working Conf. on Reverse Engineering (WCRE)}, pages
  401--408, 2013.

\bibitem{Uddin:2012:TAA}
G.~Uddin, B.~Dagenais, and M.~P. Robillard.
\newblock Temporal analysis of api usage concepts.
\newblock In {\em Proc. of the 2012 Inter. Conf. on Software Engineering}, ICSE
  2012, pages 804--814, Piscataway, NJ, USA, 2012. IEEE Press.

\bibitem{Wang:2013:MSH}
J.~Wang, Y.~Dang, H.~Zhang, K.~Chen, T.~Xie, and D.~Zhang.
\newblock Mining succinct and high-coverage api usage patterns from source
  code.
\newblock In {\em Proc. of the 10th Working Conf. on Mining Software
  Repositories}, MSR '13, pages 319--328, Piscataway, NJ, USA, 2013. IEEE
  Press.

\bibitem{Bruch:2006:FIS}
M.~Bruch, T.~Sch\"{a}fer, and M.~Mezini.
\newblock Fruit: Ide support for framework understanding.
\newblock In {\em Proc. of the 2006 OOPSLA Workshop on Eclipse Technology
  eXchange}, eclipse '06, pages 55--59, New York, NY, USA, 2006. ACM.

\end{thebibliography}

\end{document}